\documentclass[english,aps,prl,twocolumn,superscriptaddress, groupedaddress]{revtex4}
\pdfoutput=1
\usepackage[T1]{fontenc}
\usepackage[latin9]{inputenc}
\usepackage[a4paper]{geometry}
\usepackage{textcomp}
\usepackage{amstext}
\usepackage{graphicx}
\usepackage{esint}

\makeatletter

\newcommand{\lyxmathsym}[1]{\ifmmode\begingroup\def\b@ld{bold}
  \text{\ifx\math@version\b@ld\bfseries\fi#1}\endgroup\else#1\fi}

\providecommand{\tabularnewline}{\\}

\@ifundefined{textcolor}{}
{%
 \definecolor{BLACK}{gray}{0}
 \definecolor{WHITE}{gray}{1}
 \definecolor{RED}{rgb}{1,0,0}
 \definecolor{GREEN}{rgb}{0,1,0}
 \definecolor{BLUE}{rgb}{0,0,1}
 \definecolor{CYAN}{cmyk}{1,0,0,0}
 \definecolor{MAGENTA}{cmyk}{0,1,0,0}
 \definecolor{YELLOW}{cmyk}{0,0,1,0}
}

\usepackage{babel}
\makeatother

\usepackage{babel}
\begin{document}

\title{Compact Measurement Station for Low Energy Proton Beams}
\date{\today}

\author{H. Yildiz }
\affiliation{Istanbul University, Machine Technology Program, Istanbul,
TURKEY.}

\author{A. Ozbey }
\affiliation{Istanbul University, Aircraft Technology Program, Istanbul, TURKEY.}

\author{S. Oz }
\affiliation{Istanbul University, Department of Mechanical Engineering, Istanbul, TURKEY.}

\author{B. Yasatekin }
\affiliation{Ankara University, Graduate School of Natural And Applied Sciences, Ankara, TURKEY. }
\affiliation{TAEK, Saraykoy Nuclear Research and Training Center, Ankara, TURKEY.}

\author{G. Turemen}
\affiliation{Ankara University, Graduate School of Natural And Applied Sciences,
Ankara, TURKEY. }
\affiliation{TAEK, Saraykoy Nuclear Research and Training Center, Ankara, TURKEY.}

\author{S. Ogur}
\affiliation{Bogazici University, Department of Physics and Astronomy, Istanbul, TURKEY.}

\author{E. Sunar}
\affiliation{Bogazici University, Department of Physics and Astronomy, Istanbul, TURKEY.}

\author{Y. O. Aydin} 
\affiliation{FiberLAST Inc., Ankara, TURKEY.}

\author{Veliko A. Dimov }
\affiliation{CERN, Beams Department, Geneva, SWITZERLAND.}

\author{G. Unel}
\affiliation{University of California at Irvine, Department of Physics and Astronomy, Irvine, USA.}

\author{A. Alacakir}
\affiliation{TAEK, Saraykoy Nuclear Research and Training Center, Ankara, TURKEY.}

\begin{abstract}
A compact, remote controlled, cost efficient diagnostic station has
been developed to measure the charge, the profile and the emittance
for low energy proton beams. It has been installed and tested in the
proton beam line of the Project Prometheus at SANAEM of the Turkish
Atomic Energy Authority.
\end{abstract}
\maketitle

\section{Introduction}

Current experiments with ever increasing beam intensities are pushing
proton accelerators to higher beam powers. Moreover a large number
of upgrades or new high power accelerators are on the horizon. These
modern proton beam lines start with an ion source followed by a low
beta accelerator cavity, generally a radio frequency quadrupole. It
is crucial to know the beam properties right after the ion source
and just before the low beta accelerator cavity in order to optimize
the low energy beam transport (LEBT) system. The properties in question
are the beam charge, the beam profile and the beam emittance. A simple,
compact and efficient diagnostic station to measure these quantities,
is a requirement for all low energy beamlines. We have developed such a compact,
remote controlled, cost efficient diagnostic station to this end.
The diagnostic station can be inserted between the LEBT solenoids,
allowing the beam to pass though during normal operations. When a
beam related information is needed the relevant sensor device can
be moved into measurement position and data can be obtained in a completely
automated way. The remainder of the manuscript explains the requirements,
the design and the tests of this compact measurement station. It has
been installed in the proton beam line of the SANAEM Project Prometheus
(SPP) at the Turkish Atomic Energy Authority (TAEK)'s Saraykoy Nuclear
Research and Training Center (SANAEM) \cite{overall}.

\section{Requirements and Design }

\begin{figure}
\begin{centering}
\includegraphics[width=0.9\columnwidth]{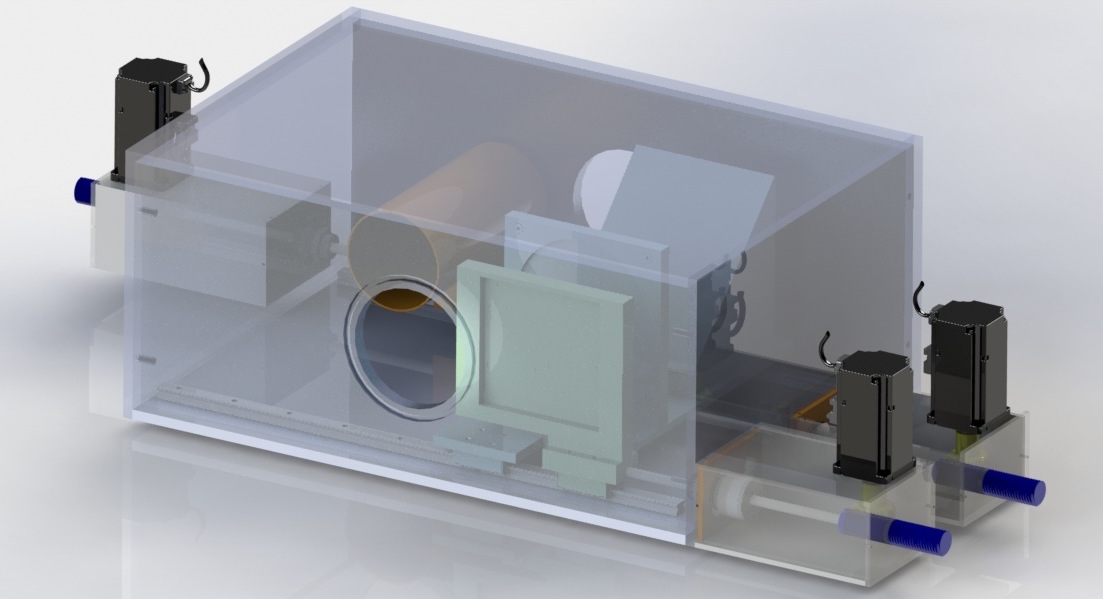}
\par\end{centering}

\caption{Compact Measurement Station Design: The beam is assumed to be moving
from left to right.\label{fig:design}}
\end{figure}

Given the previously discussed requirements, our approach to the problem
was to envisage a closed box under vacuum (at least 10$^{-6}$ mbar)
which would receive the beam, do the necessary charge, profile and
emittance measurements and provide analogue outputs. The simplicity
and efficiency requirement could be met by simply inserting the measurement
device into the beam during the beam diagnostics data collection and
by pushing it out when completed. The precise motion of the measurement
devices can be achieved by using computer controlled servo motors equipped with flanges and o-rings to preserve the vacuum.
Such a design could be used in any beam line by simply transporting
or duplicating the measurement box. A compact measurement box design
that would match these criteria  is shown in Fig.\ref{fig:design}
and its details are discussed below.

The charge measurement can be accomplished by collecting the beam
in a Faraday Cup. This charge can be transferred out of the box as
an analogue signal which either can be digitized by a QDC or can be
read by an ammeter. 

The beam profile can be obtained by intercepting the beam with an
appropriate scintillator. The scintillation light is rotated 90 degrees
by using a 45 degrees planar mirror and then collected with a digital
camera overseeing the mirror through a glass opening.

The same setup of the beam profile can be used for the emittance measurement
with an additional pepper pot plate. The pepper pot method has been
selected for its relative simplicity and cost effectiveness. It also
permits measuring emittance values in $x$ and $y$ planes simultaneously.
The associated software to convert camera output to emittance values
will be discussed in section \ref{sub:Emittance-Measurement}. After
considering the challenges of engineering and production of a detector,
we aimed to manufacture the cheapest and easiest device while trying
to keep it reliable and robust. Therefore, the Pepper Pot(PP) method
steps forward since it only necessitates a PP plate, a scintillating
view screen, and a commercially available camera or CCD, as the only
required electronics for DAQ except the computer. The challenge in
this detector will be the design, since the PP is supposed to have
a beam specific design. Also, one should underline that PP has a divergence
pre-requisite for the beam to be measured.

\begin{figure}
\begin{centering}
\includegraphics[width=0.9\columnwidth]{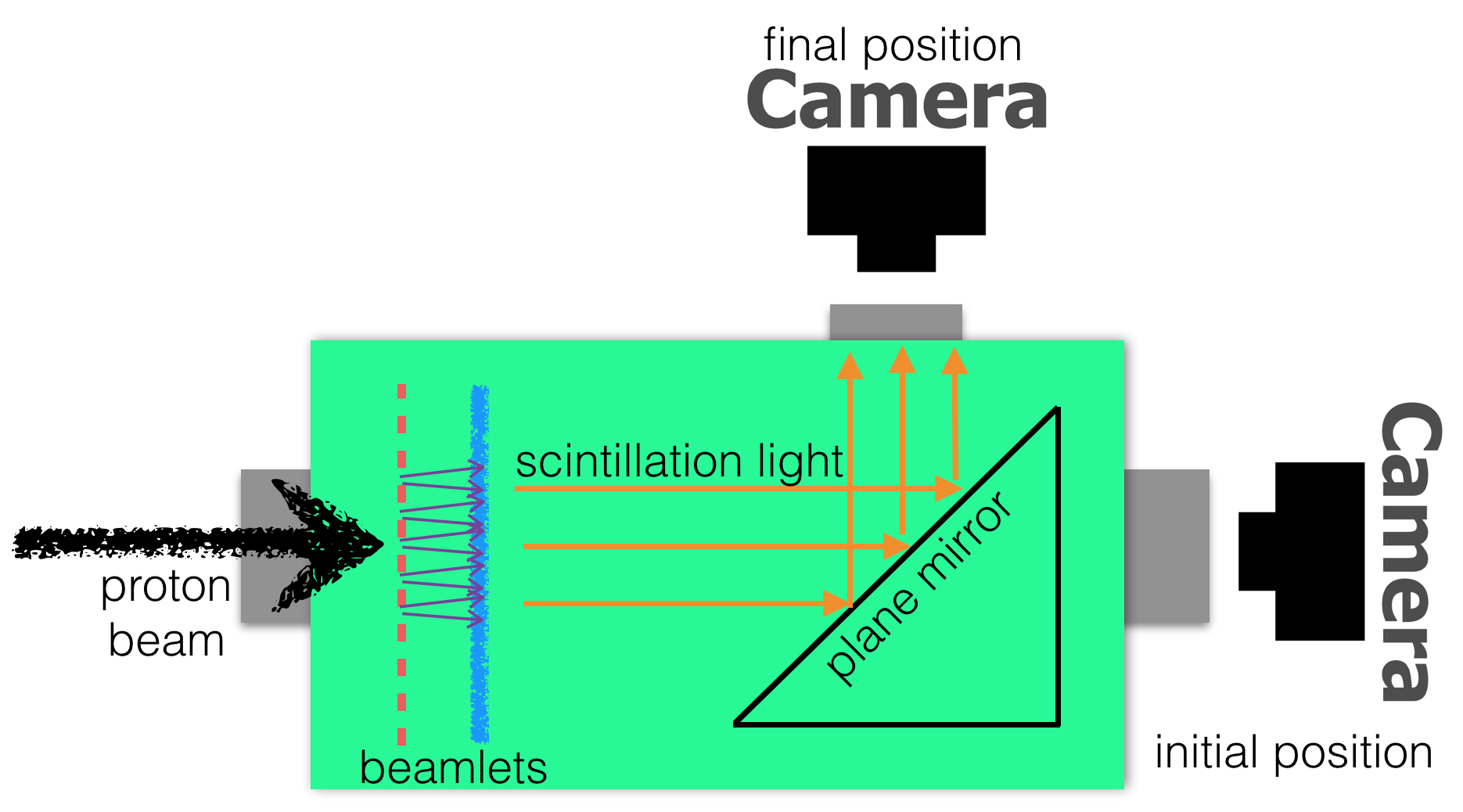}
\par\end{centering}
\caption{The positioning of the pepper pot (dashed vertical line), the scintillation
screen (rough vertical line) and the 45 degree planar mirror (triangle) inside the measurement box.
\label{fig:tuzluk-diagram}}
\end{figure}

A diagram of the emittance and profile measurement apparatus is given in Fig.\ref{fig:tuzluk-diagram}.
The initial beam moving from left to right hits the pepper pot plate shown by a dashed vertical line, producing beamlets shown by diverging arrows which in turn hit the scintillation screen shown by a rough vertical line. The scintillation light is initially read from the output end of the measurement station, and finally from the top of the box using a 45 degree planar mirror shown as a triangle.

\begin{figure}
\begin{centering}
\includegraphics[width=0.9\columnwidth]{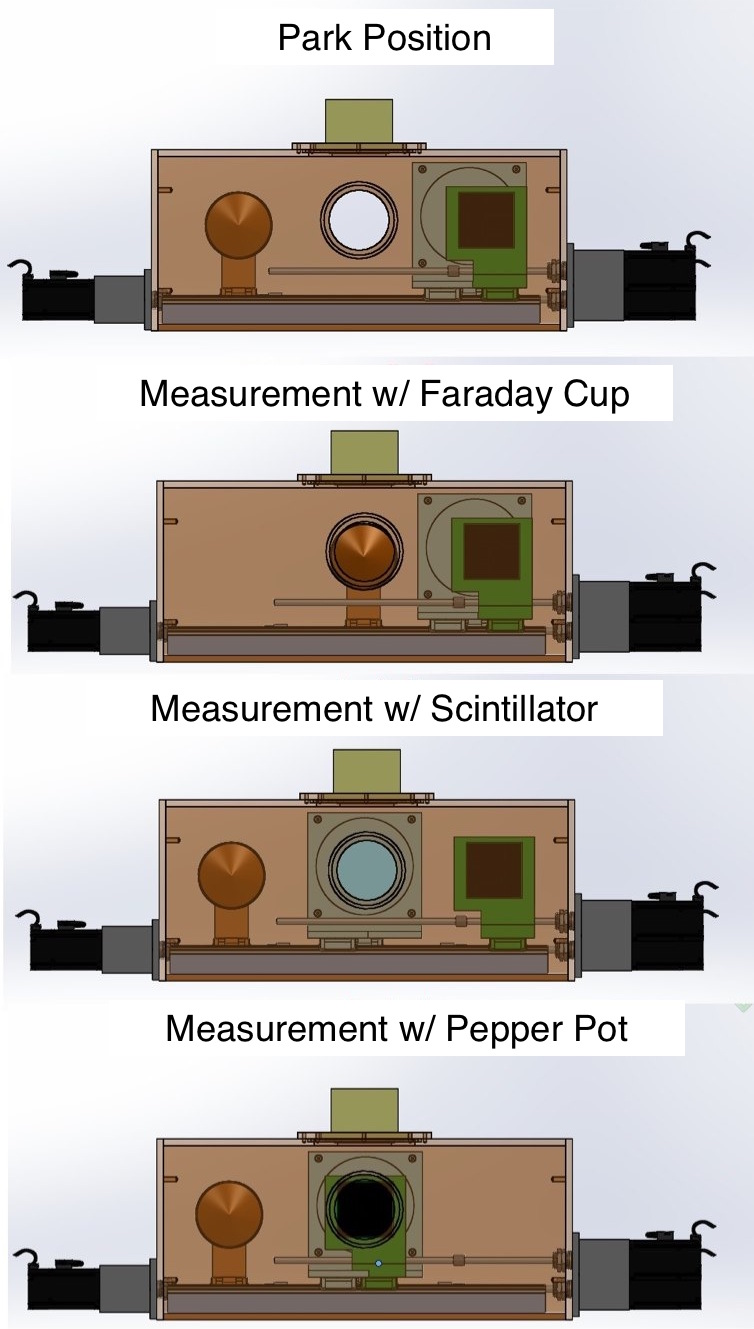}
\par\end{centering}

\caption{Compact Measurement Station in 4 different positions\label{fig:4-positions}}
\end{figure}

The diagnostic station's motion system is driven by external servo
motors with reducers. The diagnostic components like the scintillator,
PP plate, faraday cup are moved into the beam and out to the parking
position on endless screws with linear bearings. The servo motors
provide a linear accuracy of 0.4 $\mu$m. Fig \ref{fig:4-positions}
contains the sensors of the measurement box in four possible positions.
These are (listed from top to bottom) as the parking position to let
the beam go through, charge, beam profile and emittance measurement
positions.

\subsection{Mechanical Design and Manufacture}

Diagnostic Station's container box has been designed as a 380x620x395
mm rectangular box, and manufactured from (6000) Aluminum plates of
20mm thickness. The main body has been welded together, as well as
the flanges which serve as ports to other components such as vacuum
gauge, and beam pipes. The top cover is attached to the main body
by mechanical means using screws. The vacuum sealing between the body
and the top cover is achieved by using an Acrylonitrile Butadiene Rubber o-ring. The finished
diagnostics station including the servo motors, camera setup, vacuum
pump and vacuum gauge is currently installed at the SPP beamline,
between the low energy beam transport solenoids.

\subsection{Structural Analysis}

The vacuum produced in the measurement box leads a force acting on
the inner walls of the system. Bending rate of the system in vacuum
needed to be under acceptable thresholds. The vacuum behavior of the
box body made from Aluminum alloy is simulated with FEM technique
using ANSYS 14.0 \cite{ansys}. 

\begin{figure}
\begin{centering}
\includegraphics[width=0.95\columnwidth]{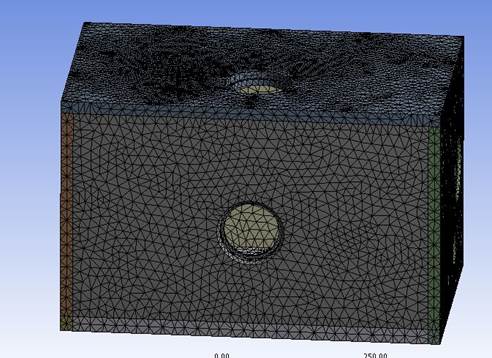}
\par\end{centering}
\caption{The mesh structure implemented in ANSYS for FEM analysis\label{fig:The-mesh-structure}}
\end{figure}

The mesh structure implemented for this analysis is shown in Fig.\ref{fig:The-mesh-structure}.
It contains 142324 elements and 237348 nodes. The element quality
has been investigated by orthogonal quality method and it is observed
that the result is above 0.3. The pressure on outer walls is 1 Atm,
while, as a result of vacuum on inner walls it is $10^{-7}$ Torr.
The body is fixed from two points at the bottom. The gravity effect
has been also included in the analysis. 

\begin{figure}
\begin{centering}
\includegraphics[width=0.95\columnwidth]{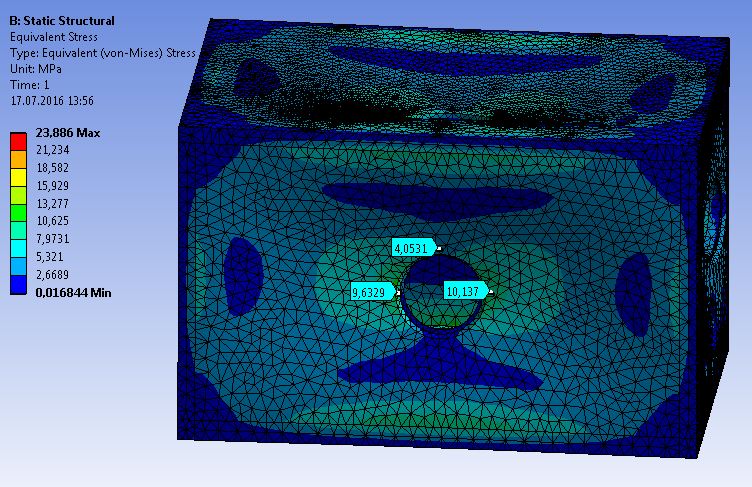} \\
\includegraphics[width=0.95\columnwidth]{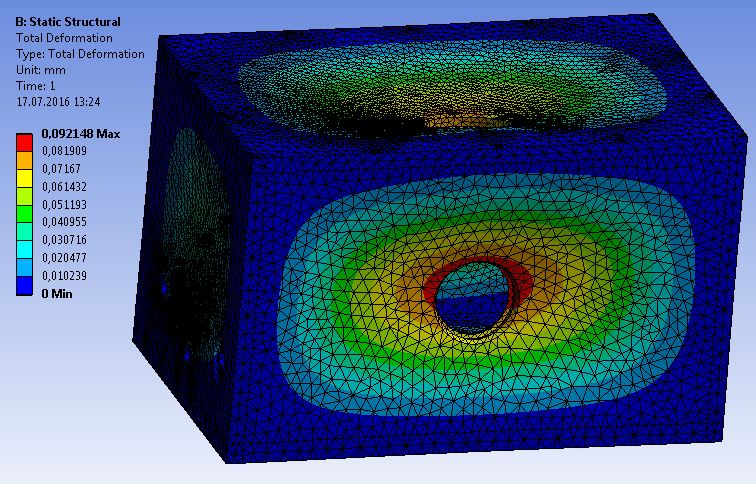}
\par\end{centering}
\caption{Up: The equivalent stress; Bottom: the total deformation both from ANSYS simulations.\label{fig:results}}
\end{figure}

According to the results of the analysis, the highest stress has been
observed in surroundings of the beam pipe which is placed in the middle
of the body. Figure \ref{fig:results} left side shows that the stress
rate is below 11 Mpa in this region. From the right side image of the
same figure it is understood that the deformation of the geometry
is less than 10 microns. The safety factor is calculated as 10.72
by using the maximum shear stress method. It is obvious that the obtained
stress and deformation rates are within acceptable ranges.

\subsection{Thermal Simulations}

\begin{figure}
\centering{}\includegraphics[width=0.9\columnwidth]{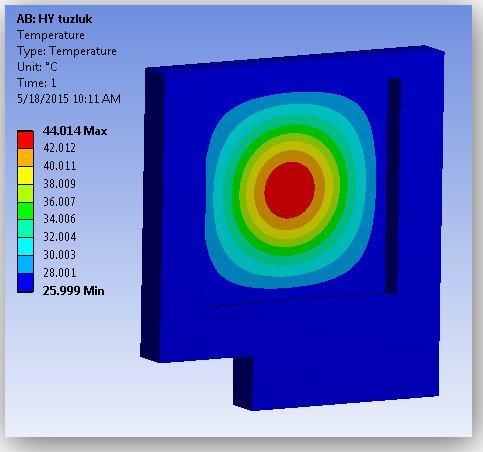} \caption{Heating for a bare square shaped Tungsten Plate with 100 $\mu$m thickness
and 10 cm sides from ANSYS simulations. \label{w_sole}}
\end{figure}

In the decision of PP plate, not only the radii of the pinholes are
crucial, but also the thickness of the plate plays a significant role.
When the protons hit the inner surface of the pinhole, they can free
secondary particles which will introduce noise, therefore, the width
of the PP plate has been determined to be as small as possible which
we conclude as 100 $\mu$m. This thickness is feasible to form and
maintain a flat shape while the passage of particles calculated by
Bethe-Bloch formula is well-above this thickness. Therefore, we only
leave with the heat and its consequences such as deformation and thermal
expansion. These two reasons forced us to consider Tungsten, Rhenium,
Tantalum kind of refractory metals. We end up to use Tungsten which
has the lowest thermal expansion coefficient. The following simulations
are made with ANSYS. If we consider the worst case scenario or in
other words, more heat-releasing situation which is when we get 20
mA of current at 20 keV of ion-source extraction within a beam size
of $\phi$=3 mm. Therefore, we have a total of 4 $W$ when the duty
factor is set to 1\%, yet we consider to take 100 seconds of measurement.
The preliminary result calculating the heat release for sole PP plate
of thickness of 100 $\mu$m with a size 10 by 10 cm is a rise of 22$^{\circ}$C
in the plate temperature as it is shown in Figure \ref{w_sole}. 
For the initial application of the measurement station, the SPP LEBT, a repetition rate
of 1Hz corresponds to a duty factor of about \%0.015. Such a low duty factor can lead to a lack of space charge neutralization. And in turn, the space charge effects may cause emittance growth in pulsed beams that may not be there in DC beams with neutralization. However, since the designed RFQ will also operate in pulsed mode, the measured emittance value is the one that would matter for this case. Additionally for the first application, the low ($\leq$1mA) current beam's space charge effects could be considered negligible for practical purposes.

\begin{figure}
\centering{}\includegraphics[width=0.9\columnwidth]{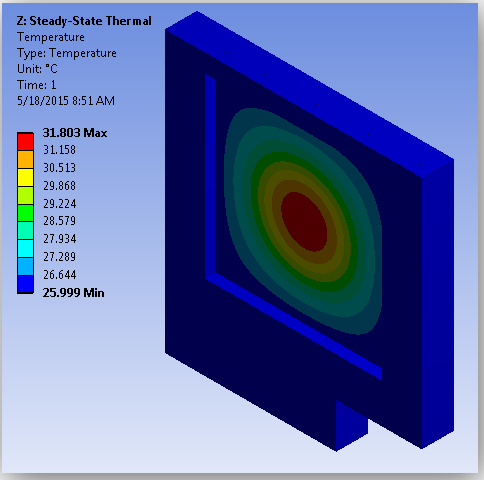} \caption{Heating for Tungsten plate supported by 0.5 mm thick Al plate and
Al rails (omitted in illustration) from ANSYS simulations. \label{comple} }
\end{figure}

Hence, we consider to support the tungsten PP plate with an Aluminum
Plate with a couple of times larger in thickness and in pinhole diameter
which will to maintain the rigidity of the frontier W plate, and to
avoid over-heating. Besides, the W and Al plates are held with Al
rails and Al frame such that the heat is conducted through the electrically
grounded and air cooled diagnostics box. The simulated maximum temperature
of complex illustrated in Fig.\ref{comple} is about 32$^{\circ}$C,
which will result in negligible lengthy thermal expansion, yet providing
fairly enough environment for fault free operation.

\section{Software Aspects}

\subsection{Control and Data Acquisition }

The observation and control of the moving parts of the diagnostic
station which has a closed structure has been done. For the realization
of this task, the following components have been added to the system. 

Electric Motor: Since a measurement system will be controlled, it
is necessary to know the position of the components continuously and
to move them to the desired positions with high precision. For this
reason, servo motors have been used on the system. By using this motors,
position and velocity of the shaft can be controlled accurately in
a very wide range of rpm without any additional component. To satisfy
the needs of high repeatability, safety and reliability of the system,
the industrial standards have been adopted. Thus, both their communication
capabilities and structures, the servo motors which fit to industrial
applications have been used. Resulting from their low maintenance
needs and more stable characteristics, the AC servo motors have been
preferred. Both the advantage of price/performance ratio and the brands\textquoteright{}
technical support capabilities at the region, Delta AC ASD-A2 servo
motor and driver have been chosen. 

Sensors: As mentioned before, servo systems can provide position information
continuously. But they need a reference point to identify their actual
position when the system turned on. Thus, the servos are provided
to move a homing position in order to reset the position data. When
they reached their homing positions, a digital signal is sent to the
control unit for each by sensors. Because of their structural simplicities
and robust characteristics, mechanical limit switches have been used.
For safety reasons, the limit switches have been adjusted to normally
closed configuration. 

Reducers: The servo motors are outside of the diagnostic station which
works in a high vacuum environment. The preventions are taken to measure
vacuum leaks between the motor shafts and the station cause to mechanical
strains. In order to tolerate the effects of the mechanical strains
on the servo motors, reducers are used. The negative effect of the
reducers to the speed of movable parts can be neglected in view of
needs of the system and the servo motors' top speed. In addition,
by using this reducers, the angular sensitivity of the servo motors
are increased. Because of their compact structure, high mechanical
efficiency and the system's mechanical needs, 1/10 of conversion rate
planetary type reducers have been chosen. 

Controller: To control positioning and activation precisely, to operate
all the control algorithms including alarm state and to communicate
with user interface and all other components, a controller structure
is needed. Siemens S7-1200 PLC is used to meet such requirements in
industrial standards.

\begin{figure}
\centering{}\includegraphics[width=0.9\columnwidth]{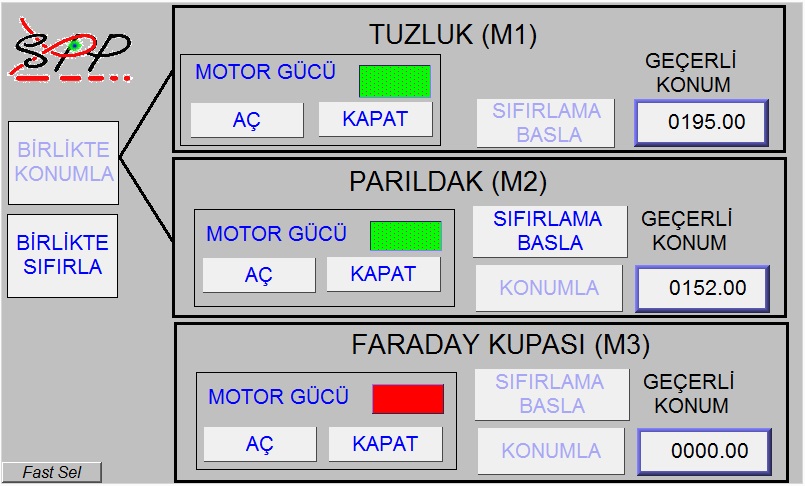}\caption{The motor GUI\label{fig:The-motor-GUI}}
\end{figure}

User Interface: In order to visualize the collected data gathered
in the PLC unit and to send the user commands to PLC unit to process
them on demand, a user interface is needed. Thus, a Weintek 8100ie
series HMI(Human Machine Interface) unit has been placed on the control
panel, which is also containing the PLC unit. The main factor of choosing
this 10\textquoteright \textquoteright{} sized touch panel is to reach
the tag based addressing data of the PLC unit directly. This provides
great convenience especially in terms of programming of the HMI unit.
A copy of the operator panel GUI, seen on Fig.\ref{fig:The-motor-GUI},
may also be operated on a personal computer connected to the system.

\begin{figure}
\centering{}\includegraphics[width=0.99\columnwidth]{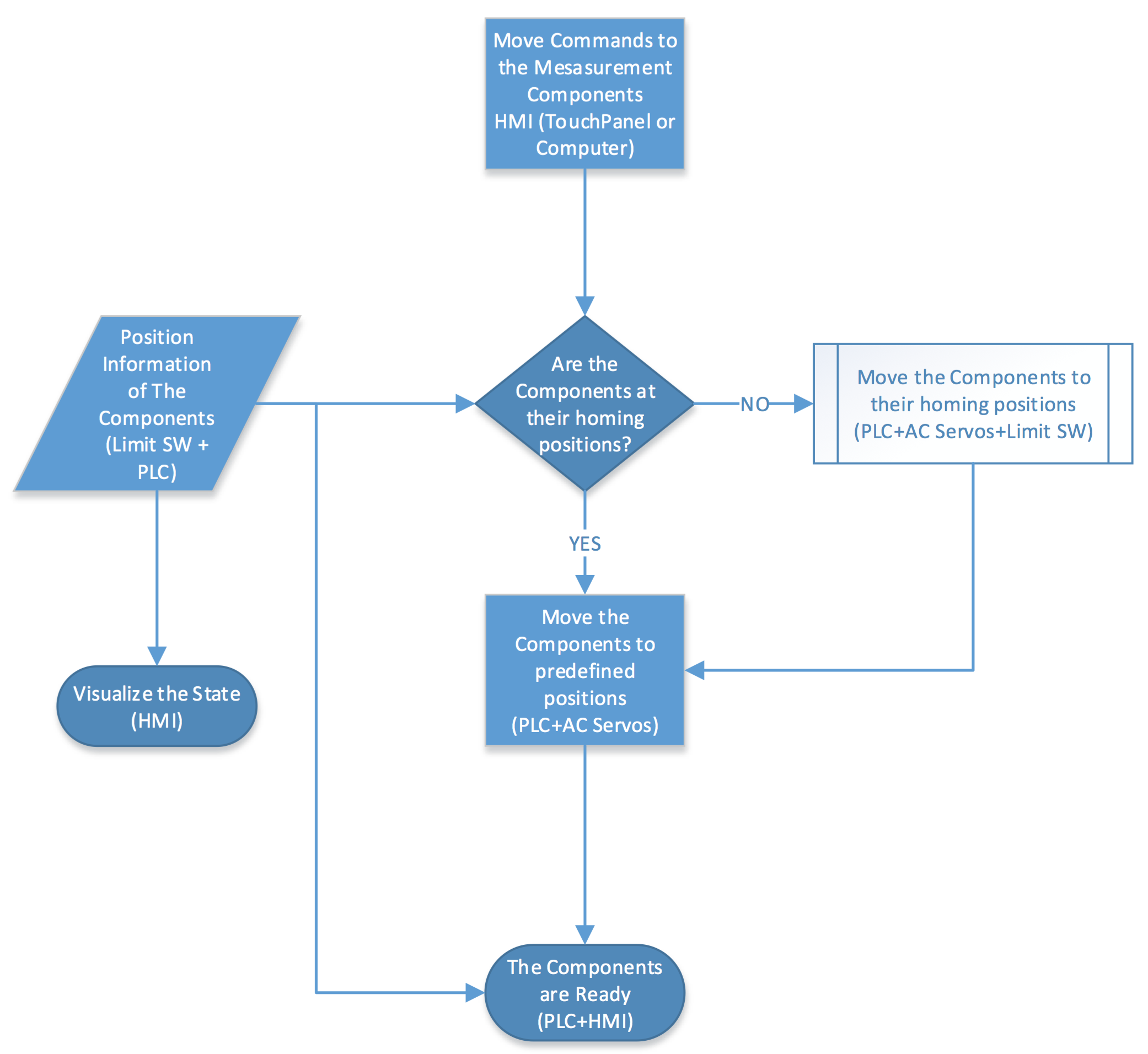}\caption{The motion control finite state machine\label{fig:FSM}}
\end{figure}

The operation of the system operation diagram is shown in Fig.\ref{fig:FSM}.
The user commands are transmitted and the current status of the system
is monitored via the HMI system which can be an operator panel and
(or) the computer screen. 

\subsection{Emittance Measurement\label{sub:Emittance-Measurement} }

\begin{figure}
\centering{}\includegraphics[width=1.05\columnwidth]{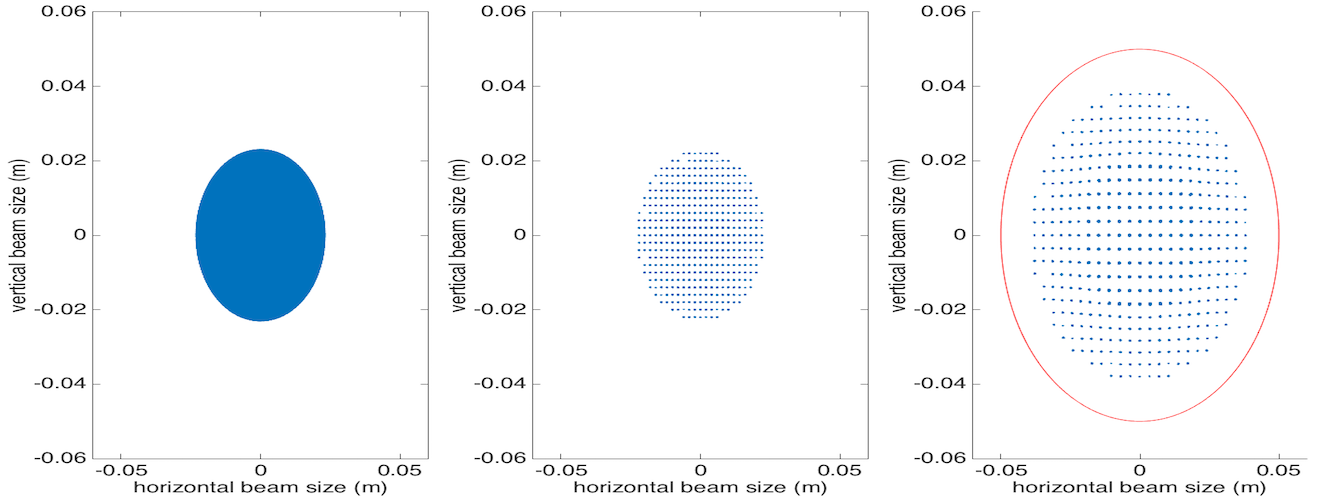} 
\caption{The leftmost is the Beam Profile, Center: beam profile just after
Pepper Pot plate, Right: drifted particles hitting the $\phi$=10cm
view screen all simulation results. \label{fig:pp} }
\end{figure}

Pepper pot (PP) emittance meter is simultaneous way of measuring both transverse
emittances. The method is similar to the work at GSI\cite{GSI} and other more recent studies\cite{PPot1, PPot2, PPot3, PPot4}. 
Briefly, the PP is a device that samples the beam through a two dimensional pinhole
grid and allows measurements of the beam parameters from the photographic images.
The validity of the algorithm and its implementation are checked using a proton beam created in IBSIMU,
an ion optical computer simulation package for ion optics, plasma extraction and space charge dominated ion beam transport using Vlasov iteration \cite{IBSIMU}. The initial proton beam profile is shown on the leftmost plot of Fig.\ref{fig:pp}. 
 The remaining particles, after sifting out both horizontally and vertically through the PP, are shown in the middle plot.
 The particles not passing through the holes are dropped from this simple tracking and simulation software written in MATLAB\cite{matlab}.
 Finally on the remaining particles are drifted along the beam direction to hit the screen, which is treated as a CCD camera for practical purposes due to the linearity of the particle motion and the proportionality of the light output to the incident beam. The expected image on the screen is shown in the same figure rightmost plot.

The reconstruction of the 4-D phase space is made by three steps.
First, the position of the particles are endorsed as the center of
the pinholes on the PP plate. Second, the beamlets create circle-like
spots on the scintillator; and the angle is determined by matching
the associated pinholes to corresponding spots. Third, the light output
assumed to be proportional to the intensity of the incident current
so that the spot density is read off as charge. All in all, a MATLAB\cite{matlab}
program has been written to simulate PP emittance meter by reconstruction
of the phase space using a multi-particle beam. In terms of design,
a solid software is an inevitable necessity to simulate how the PP
works. In one hand, since the radius of the pinhole as well as the
pinhole to pinhole distance determine the spatial resolution, it is
natural to ask pinholes to be many and tightly distributed. On the
other hand, the smaller pinholes allow fewer particles which can be
less than the cut-off current(CC) value. Besides, the closer they
are, the more likely their spots on view screen would overlap. This
phenomenon, called beamlet\textbf{ }overlap, is the most problematic
outcome of PP. We may avoid this problem up to a point by shortening
the distance between PP plate and the view screen ($L$). However,
bigger $L$ values bring about better spatial spread, conclusively
better angular\textbf{ }resolution, yet the better spread of the particles
on the screen means fewer particles incident into the unit area. This
obligates us to rethink of the beamlet current if it is below the
CC, again. This is why we have made the view screen movable on a rail
such that we avoid beamlet overlap while keeping beamlet current above
CC . Also, Monte Carlo simulations have been made to forecast possible
errors may the movement of PP plate and view screen bring. It turns
out that the horizontal movements would not have a substantial impact
whereas an error in $L$ determination together with misalignment
of PP plate and view screen cause ineligible results. 
After simulations, we conclude to manufacture the pinhole radius 50
$\lyxmathsym{\textmu}$m (or less up to 40 $\lyxmathsym{\textmu}$m)
and distance between two adjacent pinholes as 2 mm. The distance($L$)
between PP plate and view screen is adjustable between 60-150 mm to
be able to calculate emittance for our least and most divergent beam
expectations from the ion source exit\cite{ion_source}. 

\begin{figure}
\begin{centering}
\includegraphics[width=0.9\columnwidth]{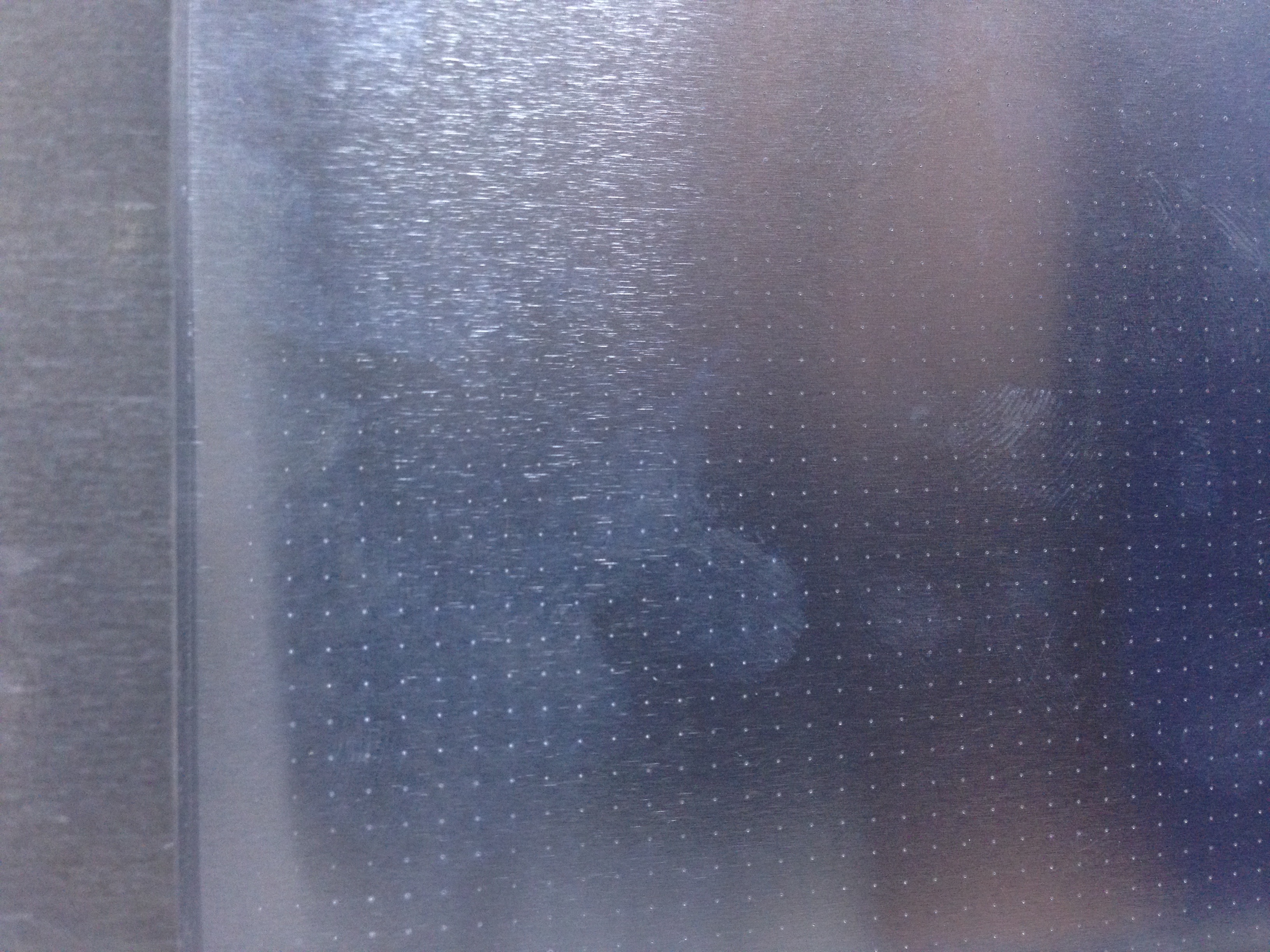}
\par\end{centering}
\caption{Pepper pot plate from aluminum\label{fig:Pepper-pot-plate}}
\end{figure}

The PP plate, is produced by a local company specializing in fiber
lasers in Ankara by drilling a thin plate with a femtosecond laser
\cite{Fiberlast}. The given design was to produce a cone like pinhole
such that the laser would aim to drill in a diameter of 110 $\mu$m
and drill out from the other side of the plate with a diameter of
about 100 $\mu$m. Early trials have been made with aluminum due of
its low cost and easiness to shape. The posterior plates are going
to be built with Tungsten. Since Tungsten has low thermal expansion
coefficient and high atomic number, these characteristics avoid thermal
deformation and result in shorter passage of protons \cite{stockli}.
The manufactured PP plate is presented in Fig.\ref{fig:Pepper-pot-plate}.
The actual hole diameters are measured to be about 80-90 $\mu$m.
This is an acceptable deviation from the design since the smaller
radii are already preferable in the PP Emittance Meter as long as
current per unit area is above the CC value. 

Apart from the simulation software, a data analysis software has been
written in MATLAB, as well. The idea is similar to PP simulation software,
yet Data Analysis code begins with reading the recorded image off
and turning it into a matrix which its size is the pixels of the image.
Since the intensity of the spot is assumed as the charge intensity,
image is read off in gray-scale. To check data analysis code, a laser
setup is prepared. A laser beam, diverged by a concave lens, is put
through the PP plate and the beamlets are incident on the tracing
paper. The image taken using an IMX145 SONY Exmor RS CMOS sensor with square pixels of width 1.5 $\mu$m \cite{sensor} 
is presented in figure \ref{fig:fp}.

\begin{center}
\begin{figure}
\centering{}\includegraphics[width=0.9\columnwidth]{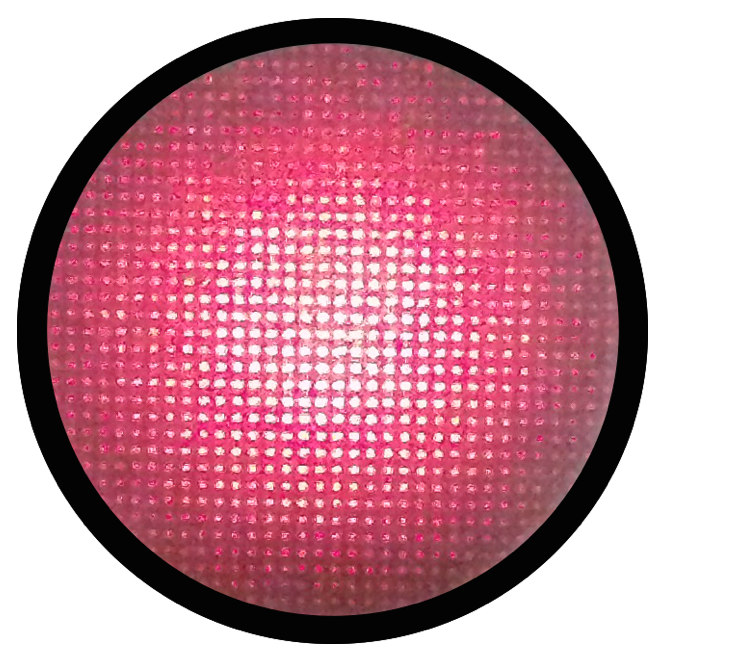} \caption{Photography recorded by a digital camera, used to check PP data analysis software. \label{fig:fp}}
\end{figure}

\par\end{center}

\begin{table}
\begin{centering}
\caption{ Changing exposure time and taking distance errors into consideration
to determine associated errors for twiss parameters. \label{table:sigma}}

\par\end{centering}

\centering{}\vspace{0.2cm}
\begin{tabular}{|c|c|c|c|}
\hline 
Exposure  & $\epsilon_{rms}[\pi.mm.mrad]$  & $\beta[mm/\pi.mrad]$  & $\alpha$ \tabularnewline
\hline 
High  & 53.0 & 3.68 & -4.62 \tabularnewline
\hline 
As is  & 51.4 & 3.91 & -4.60 \tabularnewline
\hline 
Low  & 50.8 & 3.96 & -4.20 \tabularnewline
\hline 
\end{tabular}
\end{table}

\begin{figure}
\centering{}\includegraphics[width=0.9\columnwidth]{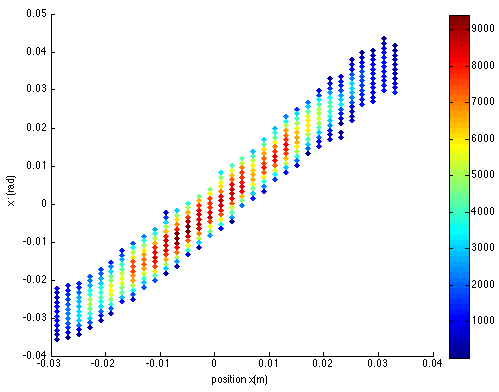}
\caption{Phase space of the laser beam diverged by the concave lens, the color
bar is constructed by the intensity of the light in each pixel processed
with 8-bits (0-255). \label{fig:masterpiece}}
\end{figure}

As a result, we conclude that the measured geometric emittance of
the laser beam is $\epsilon_{rms}=51.7\pm1.15\ \pi.mm.mrad$, and
Twiss parameters are $\beta=3.85\pm0.15\ mm/\pi.mrad$ and $\alpha=-4.47\pm0.24$.
The plot of the phase space is presented in figure \ref{fig:masterpiece}.
In fact, this calculated geometric emittance of the laser beam is
comparable to the normalized emittance of a 20 keV proton beam, $\epsilon_{n}$=$0.34\pm0.01$
$\pi.mm.mrad$, expected value at the TAEK-SANAEM's proton beamline.

\section{Assembly and Tests}

\begin{figure}
\begin{centering}
\includegraphics[width=0.9\columnwidth]{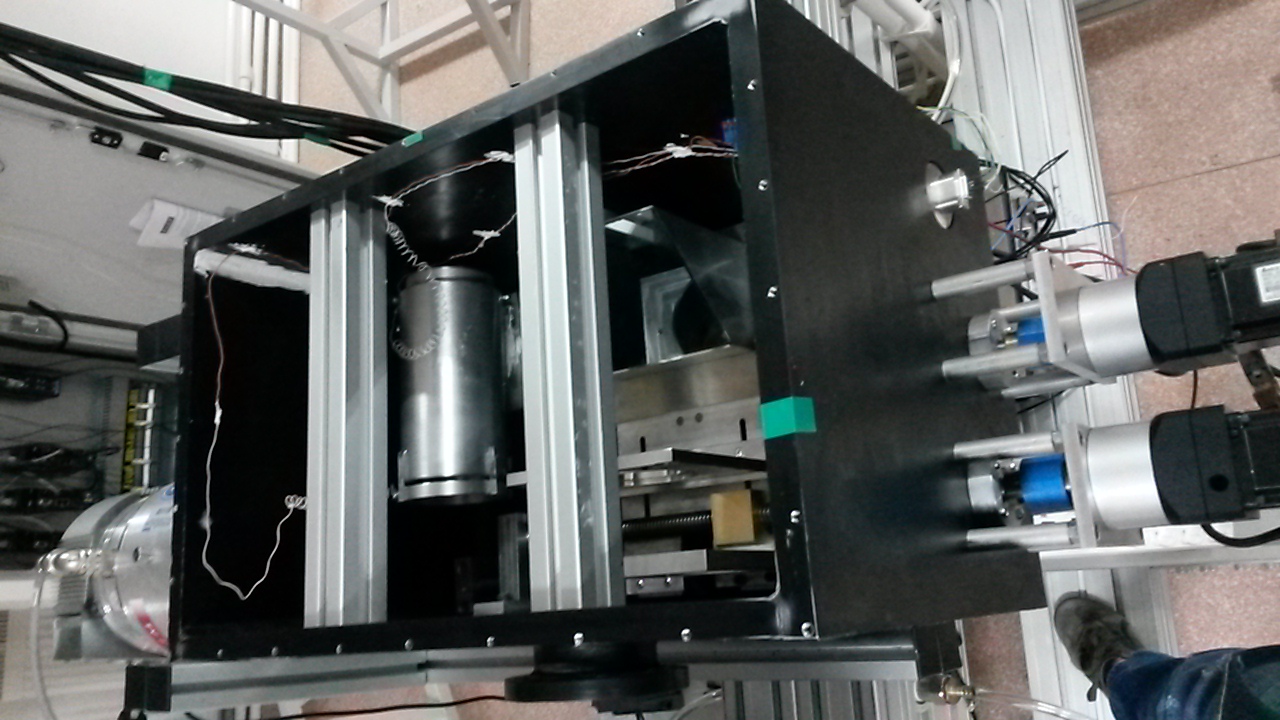}
\par\end{centering}

\caption{The compact measurement station after final assembly in the lab. The
beam moves from bottom to top. \label{fig:assembly} }

\end{figure}

The technical design drawings were submitted to the local companies
for manufacturing and initial assembly \cite{kalitek}. The final
assembly of the outer box and the motion system was completed in the SPP laboratory. 
The measurement devices such as the scintillator screen and the pepper pot were installed inside the station and laser aligned to ensure the smallest possible angle with respect to the beam axis, thus the linearity of the system response.
The scintillator material was Gd$_2$O$_2$S:Tb commonly known as P43 with a 545nm maximum emission wavelength and about 1ms decay time. 
The final product is shown in Fig.\ref{fig:assembly} just before the closing of its top
section. The beam moves from left to right, the driving motors can
be seen at the bottom of the image. After the assembly, the first
studies on the box were the vacuum tests. 

The measurement box's vacuum was initially provided by a TriScroll
(Varian TS300) mechanical pump that allowed reaching $1\times10^{-3}$
mbar \cite{triscoll}. At this pressure, vacuum leaks were searched
using a helium leak detector, Varian Leak Detector BR15 \cite{leak},
and mostly found at the joining edges of the box faces. These leaks
were repaired with Torr Seal paste, qualified for $10^{-9}$mbar vacuum.
Once the leaks were fixed, higher vacuum values were obtained using
a turbo-molecular pump (Varian Navigator 551) \cite{turbo}. The best
vacuum value, $10^{-7}$ mbar, was obtained with mechanical and turbo-molecular pumps working in tandem.
The mechanisms like rails, endless gears are than installed into the MB allowing the motion
of various detectors into and out of the beam. After initial degassing
of these elements, the vacuum value at was about $9\times10^{-6}$
mbar.

\begin{figure}
\begin{centering}
\includegraphics[width=0.9\columnwidth]{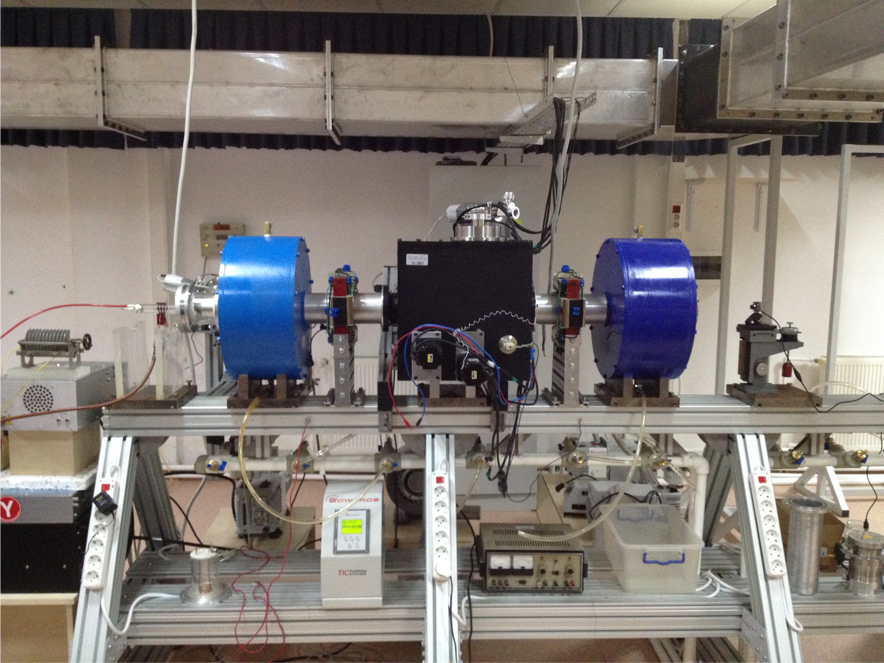}
\par\end{centering}

\caption{Measurement station installed at SPP beamline, between two blue solenoid
magnets.\label{fig:installed}}
\end{figure}

After ensuring satisfactory operation of the motors under vacuum conditions
the measurement box was installed at the low energy section of the
SPP beamline as shown in Fig.\ref{fig:installed}. At this position
i. e. after the first solenoid, the beam should be divergent, ideal
for the emittance measurement with the PP method.

\begin{figure}
\centering{}\includegraphics[width=0.9\columnwidth]{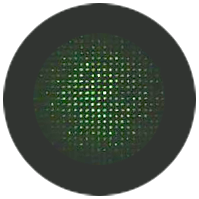}\\
\includegraphics[width=0.9\columnwidth]{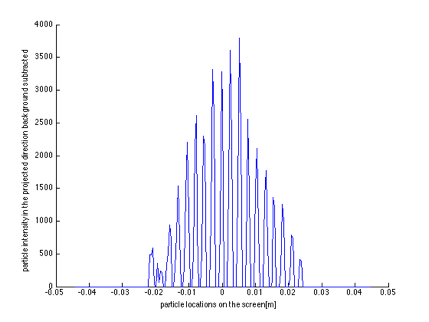}\\
\includegraphics[width=0.9\columnwidth]{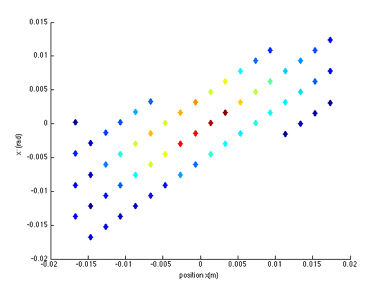}
\caption{The photography of the PP screen (top),
horizontal beam profile after the PP (middle) and the Phase space of
the 20 KeV proton beams (bottom) the color scheme is as before. \label{fig:beam-1}}
\end{figure}

A beam test was also performed to measure the emittance of the SPP
low energy proton beam. The beam profile in the horizontal direction
after the PP plate can be seen in Fig.\ref{fig:beam-1} left side
and its phase space analysis result on the same figure, right side.
The beam emittance in both horizontal and vertical directions together
with their Twiss parameters are also calculated and presented in Table
\ref{table:beam-res}. 

\begin{table}[!t]
\begin{centering}
\caption{Normalized emittance and Twiss parameter measurements with proton
beam. \label{table:beam-res}}

\par\end{centering}

\centering{}\vspace{0.2cm}
\begin{tabular}{|c|c|c|c|}
\hline 
 & $\epsilon_{rms}[\pi.mm.mrad]$  & $\beta[mm/\pi.mrad]$  & $\alpha$ \tabularnewline
\hline 
Horizontal & 0.1757 & 2.230 & -1.169 \tabularnewline
\hline 
Vertical  & 0.1688 & 5.920 & -6.241 \tabularnewline
\hline 
\end{tabular}
\end{table}

\section{Conclusions and Prospects}

The diagnostics station is working as expected. It can measure the
beam charge, profile and emittance. It can be used in other beamlines
as well after minor adjustments. For a next version of the station,
as an improvement, the vacuum shell could be produced in cylindrical
form to improve pressure resistance.

\section*{Acknowledgments}

\indent

This study is supported in part by the TAEK project under grant No.~A4.H4.P1.

\end{document}